\documentstyle[12pt,epsfig]{article}
\newcommand{\mr}[1]{{\mathrm {#1}}}

\newcommand{\ds}{\displaystyle}
\newcommand{\epem}{e$^{+}\mr{e}^{-}\;$}
\newcommand{\epeme}{\mr{e}^{+}\mr{e}^{-}}
\newcommand{\as}{$\alpha_{s}\;$}
\newcommand{\ase}{\alpha_{s}}

\newcommand{\bbar}[1]{$\overline{\mr{#1}}$}
\newcommand{\bbare}[1]{\overline{\mr{#1}}}

\newcommand{\bare}[1]{\overline{#1}}

\newcommand{\reff}[1]{(\ref{#1})}

\newcommand{\be}{\begin{equation}}
\newcommand{\ee}{\end{equation}}
\newcommand{\bd}{\begin{description}}
\newcommand{\ed}{\end{description}}
\newcommand{\bmat}{\begin{displaymath}}
\newcommand{\emat}{\end{displaymath}}
\newcommand{\bit}{\begin{itemize}}
\newcommand{\eit}{\end{itemize}}
\newcommand{\ben}{\begin{enumerate}}
\newcommand{\een}{\end{enumerate}}
\def\beq{\begin{equation}}
\def\eeq{\end{equation}}
\def\bea{\begin{eqnarray}}
\def\eea{\end{eqnarray}}
\def\bq{\begin{quote}}
\def\eq{\end{quote}}

\begin{document}
\begin{flushright}
PRA-HEP/95-03
\end{flushright}
\vspace*{0.3cm}

\begin{center}
{\huge \bf Theoretical Ambiguities of QCD Predictions at the
$Z^0$ Peak$^*$}
\end{center}

\vspace*{0.3cm}

\begin{center}
{\large
J.~Ch\'{y}la$^{a }$,
A.~L.~Kataev$^{b}$,}
\end{center}

\vspace{.5cm}

\begin{itemize}
\item[$^a$]

Institute of Physics, Academy of Sciences of the Czech
Republic, Prague

\item[$^b$]
Institute for Nuclear Research
of the Academy of Sciences of Russia,\\
117312 Moscow, Russia

\end{itemize}

\vspace{1cm}

\begin{abstract}
We discuss uncertainties of QCD predictions for the hadronic width of
the $Z^0$ boson. Emphasis is put on quantitative estimates, taking into
account the current precision of experimental data.

\end{abstract}

\vfill

\noindent
emails:
\\
chyla@cernvm.cern.ch  \\ kataev@cernvm.cern.ch
 \\

\vspace*{1cm}
\noindent
$^*$To be published in the {\em Report of the Working Group on Precision
Calculations for the Z$^0$ resonance}, CERN Yelllow Report, ed.
D. Yu. Bardin, G. Passarino and W. Hollik.

\vfill

%

\section{Introduction}
The hadronic width of the $Z^0$ boson,
$\Gamma_h\equiv\Gamma(Z^0\rightarrow \mr{hadrons})$, or the ratio
\be
R_{Z}\equiv \frac{\Gamma_h}{\Gamma_{0}},\;\;\;\;
\Gamma_{0}\equiv \frac{G_{F}M^3_{Z}}{2\pi\sqrt{2}},
\label{R}
\ee
which is closely related to the familiar ratio $R$ in \epem annihilations
into hadrons, provide theoretically very clear conditions for the
verification of perturbative QCD. This is due to several
favourable circumstances:
\bit
\item nonperturbative power corrections are expected to be
negligible at the scale $M_{Z}$
\item perturbative corrections are calculated up to
next-to-next-to-leading order (NNLO)
\item the dependence of perturbative expansion
for \reff{R} on the choice of the
renormalization scheme (RS) is weak
\item in calculating QCD corrections to \reff{R}
the running \as can be taken as that corresponding to five
effectively massless quarks, $\ase^{(5)}$
\item the effects of the finite bottom quark mass on the expansion
coefficients of $\Gamma_h$ in this couplant $\ase^{(5)}$
have been calculated up to the NLO
\item the explicit dependence of $\Gamma_h$
on the top quark mass $m_t$ has been calculated in the large $m_t$
expansion up to the NNLO
\item the statistics of the data is large,
and thus the experimental accuracy rather high,
in particular compared with
the closely related process of \epem annihilations into hadrons.
\eit
On the other hand, at the scale $M_Z$ the magnitude of QCD corrections
to the basic electroweak decay mechanism is smaller than for quantities
at lower energy scales, such as
 the $\tau$-lepton semileptonic decay width
and therefore more difficult to pin down. In this part of the paper we
give quantitiative estimates of some of the above-mentioned
uncertainties and address the related question: assuming the validity of
QCD and taking into account these uncertainties, how accurately can the
basic QCD parameter $\alpha_s$  be determined?

As any meaningful discussion of the quantitative importance of
higher order QCD corrections depends on the ability of experiments
to `see' them, we start by recalling the relevant
experimental data \cite{PDG}:
\begin{eqnarray}
M_{Z} & = & 91.187\pm0.007\; \mr{GeV}
\label{data0} \\
\Gamma_{0} & = & 0.99528\pm 0.00023\; \mr{GeV} \label{data1} \\
\Gamma_h & = & 1.7407\pm 0.0059\; \mr{GeV},
\label{data2}
\end{eqnarray}
which imply the following relative errors, relevant for further
discussion
\be
\frac{\Delta\Gamma_{0}}{\Gamma_{0}}=2.3\;10^{-4},\;\;\;
\frac{\Delta\Gamma_h}{\Gamma_h}
\doteq \frac{\Delta R_{Z}}{R_Z}=3.5\;10^{-3}.
\label{errors}
\ee
Note that the precision $\Delta\Gamma_{0}$
of the determination of $\Gamma_0$ is more
that an order of magnitude better that that of $\Gamma_h$
 and can thus be neglected with respect to
$\Delta \Gamma_h$.

Because
the dominant part of perturbative QCD predictions for $\Gamma_h$
has the same generic form as does $R_Z$, i.e.,
\be
A\left[1+\frac{\ase}{\pi}\left(1+r_1
\frac{\ase}{\pi}+\cdots\right)\right],
\label{generic}
\ee
and $\ase(M_{Z})/\pi\doteq 0.037$,
these errors allow $\ase(M_z)$ to be determined to within about 8.5\%
acccuracy.  Translated into the sensitivity to $\Lambda$, this amounts
to a factor of 1.9 uncertainty. Improving further the accuracy of the
data by a factor of two would allow it to be extracted with
an error of only 36\%.

\section{The RS dependence: general considerations}
Over the last 15 years the problem of the renormatization scheme
dependence of finite order approximants to perturbation expansions in
QCD
(and other theories) has been the subject of lively and sometimes even
heated debate. From time to time a `resolution' of this problem is
 announced, but invariably it turns out that these `solutions'
contain the original ambiguity in some guise ot another.
We intend to provide a
concise and balanced review of all the various approaches to this
problem, but emphasize at the very beginning that, in our view, there
is no clear winner. Nevertheless, as the dependence of finite order
perturbation expansions on the choice of RS is a very real
phenomenological problem, which cannot be ignored, we think the
right question in this context is: How sensitive are these
approximations to the choice of RS? But even in this question
there hides a catch,
as to give it a concrete meaning we first have to define the set of
`allowed' RS.  The point is that without some restriction on the
considered RS we could get essentially any result we want. But again, as
the selection of the `allowed' RS is inevitably a subjective matter
and may, moreover, depend on the quantity in question, the best we can
do is choose a couple of approaches which are sufficiently general,
have some rationale behind them and define the theoretical `error'
with respect to this set of RS.

Whether the theoretical error of some quantity
should be considered large or
small is, of course, not given a priori, but depends on the accuracy of
experimental data to which it is compared. These experimental
errors for the quantities related to the $Z^0$ decay were estimated in
the preceeding section.

In this section only the RS dependence of physical quantities will
be discussed. For unphysical quantities, such as the Green functions
with anomalous dimensions, the situation is more complicated and
some of the approaches are not directly applicable. This,
however, is not a serious limitation, as what we are actually
interested in are clearly only the physical quantities.

Furthermore,
 we shall consider only the case of QCD with $n_f$ massless
quark flavours. The reasons for this restrictions are twofold. First,
the relations resulting from the renormalization group (RG)
considerations and expressing the internal consistency of the
renormalized perturbation theory, have yet to be worked out for the
general massive case. The lack of such relations precludes the
general quantitative discussion of the RS dependence problem,
possible in the massless case. In the $Z^0$ mass range, however,
 we can
with great accuracy consider QCD with five effectively massless
quark flavours. This statement is quantified in Section 3.

In the following subsections we first discuss the quantitative
description of the freedom connected with the choice of the RS
(`kinematics' of the RS dependence problem) and then
briefly review several of the approaches to choosesing one of
these RS (`dynamics' of the RS problem).
We emphasize this distinction,
 as the two aspects are frequently mixed up. Only the
latter aspect is really of substance, the former being merely a matter
of convention and bookkeeping.

\subsection{The description of the RS dependence}
Consider the generic perturbation expansion for the physical quantity
of the form
\be
r(Q)= a(\mr{RS})\left[r_{0}+r_{1}(Q,\mr{RS})a(\mr{RS})+
 r_{2}(Q,\mr{RS})a^2(\mr{RS})+\cdots\right];\;\;\; r_{0}=1,
\label{r(Q)}
\ee
which appears in the expression for $\Gamma_h$ and \reff{R}.
$Q$ in \reff{r(Q)} denotes generically
some external momentum on which $r$ depends
\footnote{In view of the application to $\Gamma_h$ we restrict our
considerations to quantities depending on a single external
momentum $Q$.}
and $a(\mr{RS})$ is the {\em renormalized couplant}
$a\equiv \ase/\pi$ (the adjective renormalized will be dropped in the
following).  There are
many different ways to quantify the dependence
of such physical quantities on the RS. As a matter of
convention, we shall adopt the one suggested in Ref. \cite{PMS}.
First, we
should define the meaning of the renormalization scheme itself. In
massless QCD there must be some parameter with the dimension of the
mass, for the moment loosely denoted as $\Lambda$, that sets the basic
scale of the theory. Once this parameter is given, any quantity can in
principle be calculated as a concrete number. For a given $\Lambda$,
fixing the RS means specifying the values of all perturbative
coefficients $r_k(\mr{RS})$, as well as the value of the expansion
parameter $a(\mr{RS})$ itself.

The labelling of the RS suggested in Ref. \cite{PMS} starts with the
familiar equation
\be
\frac{\mr{d}a(\mu,\mr{RS})}{\mr{d}\ln\mu}\equiv \beta(a)=
-ba^2(\mu,\mr{RS})
\left(1+ca(\mu,\mr{RS})+c_{2}a^2(\mu,\mr{RS})+\cdots\right),
\label{RG}
\ee
expressing the dependence of $a$ on the scale $\mu$, which
inevitably appears in the theory during the process of
renormalization.  The first two coefficients on the r.h.s. of
\reff{RG}, i.e. $b,c$ are unique functions of the number
$n_f$ of massless quarks
\be
b=\frac{11N_{c}-2n_{f}}{6};\;\;\;
c=\frac{51N_{c}-19n_{f}}{11N_{c}-4n_f},
\label{bc}
\ee
but all the higher order ones are {\em completely arbitrary}.
Once they are given and some initial condition on $a$ is specified,
\reff{RG} can be solved. The way of specifying the boundary condition is
ambiguous, but its choice is a matter of convention only. One way of
doing this is via the scale parameter
$\tilde{\Lambda}$ introduced in the
following implicit equation for the solution of \reff{RG} \cite{PMS}:
\be
b\ln\frac{\mu}{\tilde{\Lambda}}=\frac{1}{a}+c\ln\frac{ca}{1+ca}+
\int^{a}_{0}dx\left[-\frac{1}{x^2 B^{(n)}(x)}+\frac{1}
{x^2 (1+cx)}\right],
\label{okraj}
\ee
where
\be
B^{(n)}(x)\equiv (1+cx+c_2 x^2 + \cdots +c_{n-1}x^{n-1}).
\label{B}
\ee
Note that as the integral on the r.h.s. of \reff{okraj} behaves like
${\cal O}(\ase)$ the higher order coefficients $c_k;k\ge 2$ have
no influence on the value of $\tilde{\Lambda}$. Note also that
the parameter $\tilde{\Lambda}$ introduced in \reff{okraj}
differs from the $\Lambda$ used in most phenomenological
analyses by  a factor close to unity:
$\Lambda=\tilde{\Lambda}\left(2c/b\right)^{c/b}$.
At the next-to-leading order (NLO) -- i.e., keeping only the first two
terms in \reff{okraj} -- the solution of \reff{okraj} is often
approximated by the
first two terms of its expansion in powers of inverse
logarithms $\ln(\mu/\Lambda)$:
\be
a(\mu/\Lambda)=\frac{1}{b\ln(\mu/\Lambda)}-\frac{c}{b^2}
\frac{\ln(\ln(\mu^2/\Lambda^2))}{\ln^2(\mu/\Lambda)}+\cdots
\label{iterativeNLO}
\ee

The dependence of the couplant $a$ on the
parameters $c_i;i\ge 2$ is determined by equations similar to
\reff{RG}  \cite{PMS}:
\be
\frac{\mr{d}a(\mu,c_{i})}{\mr{d}c_{i}} \equiv \beta_{i}=-\beta(a)
\int_{0}^{a}\frac{bx^{i+2}}{(\beta(x))^2}\mr{d}x,
\label{betai}
\ee
which are uniquely determined by the basic $\beta$--function in
\reff{RG} and thus introduce no additional ambiguity.

It is obvious that a unique
definition of $a(\mu)$ at some $\mu$ requires, as well as
the specification of the coefficients $c_i;i\ge 2$, the
specification of the boundary condition, -- i.e., for instance, the
value of $\tilde{\Lambda}$. It is convenient to introduce
the concept of the {\em renormalization convention} (RC),
which is associated with a fully defined solution of \reff{RG}:
RC$\equiv \{\tilde{\Lambda},c_i;\;i\ge 2\}$.
As, however, $\mu$ always enters this
solution in the ratio $\mu/\tilde{\Lambda}$, we can either:
\bit
\item select one of the solutions to \reff{RG}, which we call
referential renormalization convention (RRC) and vary $\mu$ only
 or
\item fix $\mu$ by identifying it with some external momentum -- for
instance $Q$ -- and vary the solution of \reff{RG}, i.e., for
fixed coefficients $c_i$ the value of $\Lambda$ instead.
\eit
Both these options are completely equivalent and it is merely a
matter of taste as to which one to use. We prefer the former.
To vary both simultaneously is legal, but obviously
redundant.

As the choice of the RRC is a matter of convention only, we cannot
associate any physical meaning to the scale  $\mu$ itself. It serves to
label the RS, but only in a given RRC. In two different RRCs the same
$\mu$ may correspond to different values of the couplant $a$ as well as
the coefficients $r_k$. We emphasize this point as in many papers the
RS is chosen by identifying $\mu$ with some `natural' physical
scale of the process, such as
 the external momentum $Q$. Although such a
natural scale can usually be identified, its mere existence
 does not help fix the arbitrary scale $\mu$, as to get a
unique RS, the RCC also has to be specified. This is usually tacitly
assumed to be the $\overline{\mr{MS}}$, but there is no
theoretical argument
for this choice, except that in this RRC the coefficients $r_k$ are
often small. If, however, the magnitude of the coefficients of the
perturbative series for physical quantities would be the criterion, we
would be naturally drawn to the effective charges approach,
described below, where they actually vanish. In other words,
because the
choice of the $\overline{\mr{MS}}$ as the RRC is merely a convention,
there is no reason to set $\mu=Q$.

The above relation \reff{okraj} allows the expresssion of
$\mu/\tilde{\Lambda}$
in terms of $a$ and $c_i$ and thus the labelling the
RS by means of the set of parameters $a,c_i;i\ge 2$.
Using this way of labelling the RS is very convenient as
there is then no need to introduce the
RRC and also no possibility of referring
to the `natural' scale to fix
the RS.

In the NNLO order -- i.e.,
taking into account also the first nonunique
coefficient $c_2$ -- we have the equation
\be
b\ln\frac{\mu}{\tilde{\Lambda}} =  \frac{1}{a}+c\ln\frac{ca}
{\sqrt{1+ca+c_2a^2}}+f(a,c_2),
\label{NNLO}
\ee
where
\begin{eqnarray}
f(a,c_2) & = & \frac{2c_2 -c^2}{d}\left(\arctan\frac{2c_2a+c}{d}
-\arctan\frac{c}{d}\right);\;\;\;d\equiv \sqrt{4c_2-c^2};\; 4c_2>c^2
\nonumber  \\
         & = & \frac{2c_2-c^2}{d}\left(\ln\left|\frac{2c_2a+c-d}
{2c_2a+c+d}\right| -\ln\left|\frac{c-d}{c+d}\right|\right);\;\;\
d\equiv \sqrt{c^2-4c_2};\;4c_2<c^2.
\label{f}
\end{eqnarray}
Its solution depends on the value of $c_2$. We distinguish three
different cases:
\bit
\item $c_2=0$ (resp. $c_i=0, i\ge 2$), defining the so
called 't Hooft RC \cite{Hooft}
\item $c_2>0$, when $\beta(a)<0$ is monotonously decreasing function
of $a$ and the situation is therefore qualitatively the same as for
$c_2=0$
\item $c_2 < 0$, when $\beta(a)$ has the {\em infrared fixed point}
at $a^{*}(c_2)$, given by the equation $\beta[a^*(c_2)]=0$.
The corresponding
solution of \reff{RG} then approaches {\em finite} value at $\mu=0$
and consequently
\be
\lim_{Q\rightarrow 0}r^{(3)}(Q)=
a^*(c_{2})\left(1+r_{1}(\mu=Q)a^{*}(c_{2})+r_{2}(c_{2},\mu=Q)
a^{*2}(c_{2}) +\cdots \right)
\label{IR}
\ee
 has a finite
{\em infrared limit} at the NNLO. This case is discussed in detail in
Refs. \cite{my,MatSt}.
\eit
Note that the possibility of an infrared stable limit of finite order
approximants, which starts at the NNLO, does not have to
survive the incorporation of still higher order corrections and
its physical relevance is therefore questionable.

While the explicit dependence of the couplant on $c_i$ is given in
\reff{NNLO} and \reff{f}, the dependence of the coefficients $r_k$ on
them is determined by the requirements of internal consistency of the
perturbation theory. They imply that any finite order approximant
\be
r^{(N)}(Q)\equiv \sum_{k=0}^{N-1}r_{k}a^{k+1}
={\cal F}(\mu,c_{i},\rho_{i};i\le N-1),
\label{kon}
\ee
must satisfy the following consistency conditions:
\be
\frac{\mr{d}r^{(N)}}{\mr{d}\ln \mu}={\cal O}(a^{N+1}),\;\;\;\;\;
\frac{\mr{d}r^{(N)}}{\mr{d}c_{i}}={\cal O}(a^{N+1}).
\label{konpod}
\ee
Iterating these equations we find:
\begin{eqnarray}
r_{1}(Q/\mu) & = & b\ln\frac{\mu}{Q}+r_{1}(\mu=Q)=b\ln\frac{\mu}
{\tilde{\Lambda}}-\rho(Q/\tilde{\Lambda})  \nonumber \\
r_{2}(Q/\mu,c_2) & = & \rho_{2}-c_{2}+r_{1}^{2}+cr_{1},
\label{r1r2}
\end{eqnarray}
and similarly for still higher orders. In the above relations the
quantities $\rho, \rho_2$ etc., are RG invariants --
 i.e., contrary to the
coefficients $r_k$, they are independent of the choice of the RS.
Note that all the dependence of the perturbative approximants on
$Q$ comes exclusively through the invariant $\rho(Q/\Lambda)$,
 which can be written as
\be
\rho=b\ln(Q/\tilde{\Lambda}_{\mr{RRC}})-r_1(\mu=Q,\mr{RRC}),
\label{rho}
\ee
where the apparent dependence on the chosen RRC actually cancels
between the two terms in \reff{rho}.

A nontrivial part of any perturbative calculation boils
down to the evaluation of these invariants, the rest being essentially
a straightforward exploitation of the RG considerations based on
\reff{konpod}.
Substituting for the term $b\ln (\mu/\Lambda)$
in $r_1$ the expression \reff{okraj}, using \reff{r1r2}
 and inserting the resulting $r_k$
into \reff{kon}, any finite order approximant $r^{(N)}(Q)$ can be
expressed as an explicit function of the parameters specifying the RS,
i.e., $a,c_i;i\ge 2$, and the invariants $\rho_i$:
\be
r^{(N)}(Q)=f(\rho_j,j<N-1;a,c_i,i\le N-1).
\label{ff}
\ee

In this
representation the RS dependence of NLO and NNLO approximants is
quantitatively described by one- and two-dimensional manifolds,
respectively, and the problem of choosing the RS is equivalent
 to selecting one
particular point on these manifolds. Considered as a geometrical
exercise we identify certain special points on these
manifolds, corresponding to stationary points, where the variation
of the approximants with respect to the free parameters
vanishes locally.

In the next subsection we shall briefly describe some of the criteria
for choosing the RS, which will define the set of RS for which
we shall later estimate the theoretical uncertainty of
perturbative calculations of the quantities of interest.
We shall discuss in some more detail the approaches described
in subsections 2.5 and 2.6, as there have recently been some
new developments in them.

\subsection{Fixed RS calculations}
Because of the computational simplicity and explicit gauge invarince,
all the multiloop calculations are nowadays done
using the dimensional regularization technique. Within this technique
the $\overline{\mr{MS}}$ renormalization prescription\footnote{By
`prescription' we mean the specification of the
scale $\mu$ (by identifying it with some natural scale $Q$)
as well as of
finite parts of all counterterms necessary to cancel the UV
divergencies. Specifying the prescription implies the specification of
the RS in the above-defined sense, but not vice versa.}
is often preferred on the grounds that it
absorbs in the definition of the renormalized couplant the terms
proportional to $\ln 4\pi -\gamma_{E}$, which are considered to be
artefacts of the dimensional regularization technique. In our way of
labelling the RS, $\overline{\mr{MS}}$ corresponds
to definite values of all the coefficients
$r_k, c_k$ and a fixed, but numerically undetermined, value of $a$, which
must be extracted from comparison with experimental data.
This choice of the RS is very commonly used in phenomenological
analyses, but there is no obvious reason why it should be
preferred to, for instance, the MOM-like RS or any of the choices
discussed in the following subsections. In geometrical terms this is
reflected in the fact that the corresponding point on the hypersurfaces
defined in \reff{ff} occupies no special position.

\subsection{Principle of Minimal sensitivity (PMS)}
In this approach, suggested in Ref. \cite{PMS},
 the RS is fixed by demanding that
 \be
\frac{\mr{d}r^{(N)}}{\mr{d}a}=
\frac{\mr{d}r^{(N)}}{\mr{d}c_{i}}=0,
\label{stac}
\ee
i.e., the N-th order partial sum
has locally the property
that the full expansion must satisfy globally.
 Though there is in general no quarantee
that such a stationary point is unique or exists at all, in practical
applications to lowest order QCD quantities it works. At the NLO, when
only $a$ labels the RS, \reff{stac} reduces to
\be
2-2\rho a+2ca\ln\frac{ca}{1+ca}+ca\left(\frac{ca}{1+ca}\right)=0,
\label{stac2}
\ee
and its solution  has the form $a_{\mr{PMS}}=(1/\rho)[1+{\cal
O}(ca_{\mr{PMS}})]$.

At the NNLO we have two coupled equations for derivatives
of $r^{(2)}(a,c_2)$ with respect
to $a$ and $c_2$, which must be solved numerically. For the quantity
\reff{r(Q)} such a stationary point exists for any $\rho$ if
$\rho_2<0$ and for $\rho >\rho_{\mr{min}}(\rho_2)$ if $\rho_2>0$
\cite{my}.

Note that, contrary to the fixed RS approach, the PMS selects the RS
which depends on the type and kinematics of the process under study.
The same holds for the methods discussed in the next three
subsections.

\subsection{The method of effective charges (ECH)}
The basic idea of this approach \cite{ECH} is to choose the RS
is such a way that the relation between the physical quantity and the
couplant is the simplest possible one. For the quantity \reff{r(Q)} it
means:
\be
r(Q)=a_{\mr{ECH}}.
\label{FAC}
\ee
In this approach there is no problem with the convergence of the
perturbation expansion \reff{r(Q)} itself, but it reappears in the
perturbation expansion of the corresponding $\beta$--function (see
below).

The conditions under which the parameters $a,c_i$, or $\mu,c_i$,
can be chosen in such a way that \reff{FAC} holds can be read
directly off the consistency conditions \reff{konpod}. At the NLO,
\reff{FAC} implies the following equation for $a^{\mr{ECH}}$
\be
\frac{1}{a}+ca\ln\frac{ca}{1+ca}=\rho,
\label{aECH}
\ee
which has a solution
$a_{\mr{ECH}}=(1/\rho)(1+{\cal O}(ca_{\mr{ECH}}))$,
differing from $a^{\mr{PMS}}$ merely by the term of the order
${\cal O}(ca_{\mr{ECH}})$.  At the NLO the values of $a_{\mr{ECH}}$
thus correspond to intersections of the curves defined as
\be
r^{(2)}(a,c_2)=a\left(2-\rho a +ca\ln\frac{ca}{1+ca}\right),
\label{rn2}
\ee
with the straight line
$r^{(2)}=a$. For $\rho>0$ there is always just one such intersection
in the physically relevant range $a>0$, while for $\rho<0$ there is
none in this range.

At the NNLO the situation is more complicated, as the condition
\reff{FAC} does not by itself determine uniquely $a_{\mr{ECH}}$, but
merely implies  the equation
\be
    r_1(a)+r_2(a,c_2)a=0,
\label{allFAC}
\ee
which has, depending on the value of $c_2$,
either no solution, or one or two solutions,
giving {\em different} $a_{\mr{ECH}}$.
Going to still higher orders the
ambiguity of this approach grows even further as new free parameters,
$c_i$, crop up. In \cite{ECH} this ambiguity is avoided by
demanding that each of the coefficients $r_i$ vanishes individually.
Assuming this restricted version of the ECH method -- i.e., demanding
$r_i=0,i\ge 1$ -- we get the following expression for the associated
$\beta$--function:
\be
\frac{da_{\mr{ECH}}}{d\ln\mu}\equiv
\beta_{\mr{ECH}}(a)=-ba_{\mr{ECH}}^2
\left[1+\rho_1 a_{\mr{ECH}}+
\rho_2a_{\mr{ECH}}^2+\cdots\right],
\label{betaECH}
\ee
where $\rho_1\equiv c$. The coefficients $c_{\mr{ECH},i}$ of
the ECH $\beta$--function $\beta_{\mr{ECH}}$ thus concide with the RG
invariants $\rho_i$, introduced in \reff{r1r2}.
To express $a_{\mr{ECH}}$
as a function of the external momentum $Q$, it
is convenient to write it as a solution of the equation
\be
\frac{1}{a_{\mr{ECH}}}+c\ln\frac{ca_{\mr{ECH}}}{1+ca_{\mr{ECH}}}
=b\ln \frac{Q}{\Lambda_{\mr{ECH}}},
\label{excplic}
\ee
where $\Lambda_{\mr{ECH}}$ defines the `effective' $\Lambda$
parameter, associated with the quantity under study. It is related to
$\Lambda_{\mr{RS}}$ in any fixed RS simply as
\be
\Lambda_{\mr{ECH}}=\Lambda_{\mr{RS}}\exp\left(r_1(\mu=Q,\mr{RS})
/b\right).
\label{Leff}
\ee

As the ECH approach seems to offer a very simple and natural
`solution' to the RS problem one might naturally ask: where has all
the ambiguity discussed in subsection 2.1 actually gone? In fact
it has not disappeared entirely and reemerges, as discussed in Ref.
\cite{JaECH}, in a somewhat disguised form, even within the ECH
approach.

\subsection{The method of Brodsky, Lepage and MacKenzie}
This method \cite{BLM} borrows its basic idea from QED, where the
renormalized electric charge is fully given by the vacuum polarization
due to charged fermion--antifermion pairs. In QCD the authors
of this method suggest fixing the scale
$\mu$ with the requirement that all the effects of quark pairs be
absorbed in the definition of the renormalized couplant itself,
leaving nothing in the expansion coefficients.  In the case of the
quantity \reff{r(Q)} and up to the NNLO,
\be
r(Q)=a(\mu,\mr{RRC})\left(1+
\left[r_{10}\left(\frac{\mu}{Q},\mr{RRC}\right)+
n_f r_{11}\left(\frac{\mu}{Q},\mr{RRC}\right)\right]
a(\mu,\mr{RRC})\right),
\label{AB}
\ee
where
we have now written the $n_f$ dependence of the coefficients $r_k$
explicitly, it amounts to the requirement that $\mu$ be chosen in such
a way that $r_{11}(\mu/Q,\mr{RRC})=0$.
The problem with this `scale-setting' method is that the resulting
scale as well as $r^{\mr{BLM}}$ depend on the choice of the RRC! The is
 due
to the fact that for a given $\mu$ the separation of the coefficient
$r_1$ into the two parts $A+n_f B$ {\em is not unique}, but depends on
the RRC used.  This problem exists in principle QED as well,
but there the quark loop effects can be rather unambiguously absorbed
in the renormalized electric charge via the MOM RRC, which for
massless quarks gives the same $B$ as the MS-like ones. This is no
longer true in QCD, where various types of MOM-like RRC in
general give different values of $B$, different again from that of
$\overline{\mr{MS}}$-like one \cite{PMSC}.

As emphasized in the general discussion above, fixing the scale without
also
simultaneously fixing the RRC does not, however, determine the RS,
because
the choice of the RRC is equally important as that of the scale $\mu$.
The resulting ambiguity of the BLM approach, pointed out a long time ago
\cite{PMSC}, has not yet been satisfactorily resolved.
In practical applications  one usually
starts from the $\overline{\mr{MS}}$ RRC.

It is claimed in Ref. \cite{BLM} that
the BLM-improved expressions for the physical quantities have small
NLO coefficients.
However, as shown in Ref. \cite{GK}, this is not
necessarily the case when the BLM approach is generalized to higher
orders. Let us first recall the main steps of the generalization
suggested in Ref. \cite{GK}.

Within the class of the co called `regular' RC the $n_f$ dependence
of the expansion as well as $\beta$--function coefficients is
polynomial in $n_f$:
\begin{eqnarray}
r_1 & = & r_{10}+r_{11}n_f \label{r1nf} \nonumber \\
r_2 & = & r_{20}+r_{21}n_f+r_{22}n_f^2  \label{r2nf} \nonumber \\
\beta_0 & \equiv & b=b_{00}+b_{01}n_f \nonumber \\
\beta_1 & \equiv & bc= \beta_{10}+\beta_{11}n_f  \label{poly} \\
\beta_2 & \equiv &
bc_2=\beta_{20}+\beta_{21}n_f+\beta_{22}n_f^2+\beta_{23}n_f^3
\nonumber \\
\tilde{\beta}_2 &\equiv &
b\rho_2=\tilde{\beta}_{20}+\tilde{\beta}_{21}n_f+\tilde{\beta}_{22}n_f^2
+\tilde{\beta}_{23}n_f^3. \nonumber
\label{betanf}
\end{eqnarray}
Note that the scheme-invariant coefficient $\tilde{\beta}_2$
contains the $n_f^3$ term as observed in  Ref. \cite{Kat}.

The generalization of the BLM approach suggested in \cite{GK}
assumes that the chosen scale $\mu$ is determined by the following
perturbative expansion:
\begin{equation}
\mu^2=\mu_{\mr{BLM}}^2
\bigg(1+\gamma_1(n_f)a(\mu_{\mr{BLM}})+...\bigg),
\label{scale}
\end{equation}
where
$\mu_{\mr{BLM}}^2$ is given in
\cite{BLM}
and
$\gamma_1=\gamma_{10}+\gamma_{11}n_f$.
The parameters $\gamma_{10}$ and $\gamma_{11}$ are process
dependent and can be determined from the following system of
equations\footnote{Note the factor-of-two difference between our
definition
\reff{RG} of the $\beta$--function and that used in Ref. \cite{GK}.}
\begin{eqnarray}
\tilde{\beta}_{23}-\beta_{23}&=&-2\beta_{01}^2\gamma_{11} \nonumber \\
\tilde{\beta}_{22}-\beta_{22}&=&-2\beta_{01}^2\gamma_{10}
-4\beta_{00}\beta_{11}\gamma_{11} \nonumber \\
\tilde{\beta}_{21}-\beta_{21} &=& \beta_{01}\bigg(r_2^*-r_1^{*2}\bigg)
-\beta_{11}r_1^*-
4\beta_{00}\beta_{01}\gamma_{10}-2\beta_{00}^2\gamma_{11}^2
\nonumber \\
\tilde{\beta}_{20}-\beta_{20} &=& \beta_{00}\bigg(r_2^*-r_1^{*2}\bigg)
-\beta_{10}r_1^*-2\beta_{00}^2\gamma_{10},
\label{syst}
\end{eqnarray}
which follows from the general expression for the scheme invariant
$\tilde{\beta}_2$:
\be
\tilde{\beta}_2=\beta_2+br_2-\beta_1r_1-br_1^2.
\label{betatilde}
\ee
In the above equations,
$r_1^*$ and $r_2^*$ are the $n_f$-independent coefficients in the
generalized BLM procedure. We have already mentioned that in practice
the
BLM approach is applied to the initial series with the coefficients
defined in the $\overline{\mr{MS}}$-scheme.
Therefore, it is necessary to put $\beta_{23}=0$.
Now consider as an example perturbation expansion for the familiar
quantity:
\be
R(s)\equiv \frac{\sigma(\epeme\rightarrow \mr{hadrons})}
{\sigma(\epeme \rightarrow\mu^+\mu^-)}=
\left(3\sum_{i=1}^{n_f}Q_{i}^2\right)\left[1+\sum_{k=0}^{\infty}
r_ka^{k+1}\right].
\label{Repem}
\ee
Applying the above generalization to \reff{Repem} we get \cite{GK}
\be
\mu_{\mr{BLM}}^2=
\mu^2_{\overline{\mr{MS}}}\exp(0.69),\;\;\; \gamma_{01}=0.11,
\;\;\;\gamma_{11}\approx 3,
\label{ex}
\ee
which implies
\be
R(s)  = \left(3 \sum_{i=1}^{n_f} Q^{2}_{i}\right)\left[1 +
a_{*} +  0.08 a_{*}^2-23.3a_{*}^3\right] -\bigg(\sum_{i=1}^{n_f}
Q_{i}\bigg)^2 1.24a_{*}^3,
\label{BLMres1}
\ee
where
\be
a_{*}=a\bigg(\mu^2_{\mr{BLM}}(1+\gamma_1(n_f)a(\mu^2_{\mr{BLM}})
\bigg).
\label{astar}
\ee
Notice that the coefficient of the NNLO correction is, indeed,
not small.
Therefore it is not true, as conjectured in Ref. \cite{BLM}, that the
BLM-improved perturbative series have in general significantly
smaller coefficients than the expansions in the $\overline{\mr{MS}}$.

\subsection{RS invariant perturbation theory}
The basic problem of RS dependence can be traced back to the fact that
the expansion parameter, the renormalized couplant $a$, is not a
physical quantity, but rather an intermediate variable, allowing us
to correlate different physical quantities. As such, it is inevitably
ambiguous. This problem can be circumvented -- at least in part -- by
expressing one physical quantity directly as power expansion in terms of
the other. Consider, for instance, two physical quantities admitting the
following perturbation expansions in some
RS\footnote
{There is no problem generalizing this analysis to the case
of different powers of the leading terms.}
\begin{eqnarray} R^{(1)} & = &
a(\mr{RS})\left(1+r_1^{(1)}(\mr{RS})a(\mr{RS})+ \cdots\right)
\nonumber \\
R^{(2)} & = & a(\mr{RS})\left(1+r_1^{(2)}(\mr{RS})a(\mr{RS})+
\cdots\right)
\label{R1R2}
\end{eqnarray}
Expressing $a(\mr{RS})$ from the first in terms of $R^{(1)}$ and
substituting into the second equation we get
\be
R^{(2)}=R^{(1)}\left(1+\Delta^{(2,1)}R^{(1)}+\cdots\right),
\label{R2(R1)}
\ee
where $\Delta^{(2,1)}=r^{(2)}_1(\mr{RS}) -r^{(1)}_1(\mr{RS})$,
as well as all the other coefficients $\Delta^{(2,j)}, j\ge 2$ of
this expansion, {\em are unique}. This kind of expansion has
already been discussed within the
so-called `scheme invariant perturbation theory'
in Ref. \cite{Dhar} and recently resurected within the
so called `commensurate scale relations' in \cite{Lu}.
The essence, however, remains the same.

As a special and interesting example of such a relation, consider the
one between the derivative of \reff{r(Q)} with respect to the
external momentum $Q$ and the quantity $r(Q)$ itself. It is
straightforward exercise to show that this relation reads:
\be
\frac{\mr{d}r(Q)}{\mr{d}\ln Q}=
-b[r(Q)]^2\left(1+cr(Q)+\rho_2[r(Q)]^2\cdots
\right)=\beta_{\mr{ECH}}(r(Q)),
\label{dr(Q)}
\ee
where the r.h.s. is nothing else than the `effective'
$\beta$--function introduced above, evaluated at $r(Q)$!

In \reff{R2(R1)} as well as in \reff{dr(Q)} there is no trace of
any RS
ambiguity. It is nice to be able to show explicitly,
as is done in Refs. [12--14],
that
all the coefficients in these expansions are, indeed, RS
invariants, but it cannot be otherwise, as they relate two
physical variables and there is no way their eventual
dependence on the RS could be cancelled. Perturbation theory serves
here merely as an intermediate, but vital, tool for evaluating
coefficients such as $\Delta^{(2,j)}$ and $\rho_i$.

If relations such as \reff{R2(R1)} or \reff{dr(Q)} are truly unique,
do they not solve the whole RS--dependence problem? The answer
depends on what we expect from the perturbation theory. If we are
interested merely in relating pairs of physical quantities admitting
purely perturbative expansions the answer is positive.
If, however, we ask the question: what is, on the basis of analyses of
available experimental data, the QCD prediction for, say, the $Z^0$
width, then the answer is definitely negative. The point is that in
predicting the value of $\Gamma_h$ from the relations
between this quantity and some other physical quantity $R$,
we find that the resulting predictions {\em depend on the choice of
the quantity $R$}! What has been thrown out the door
in the form of the RS
ambiguity, comes back through the window as the `initial condition'
ambiguity \cite{JaDhar}. Moreover, this new one is even
more difficult to handle than the original one!

We therefore believe that the better way of incorporating the
results of QCD
analyses of many different physical quantities measured in different
kinematical ranges is to introduce some intermediate,
no-nunique and thus
{\em unphysical}, variable, which can then be used for QCD predictions
of other physical
quantities. The renormalized couplant serves just this purpose.

\section{Quark mass thresholds in the running \as}
The proper treatment of quark mass effects in the QCD running coupling
constant $\alpha_s$ has so far not been in the forefront of interest of
theorists and phenomenologists. This is now changing. Although the quark
mass effects are not large, steady improvement in the precision of
experimental data, coming in particular from a new generation of
experiments at CERN and Fermilab, combined with significant progress in
higher order QCD calculations, has led to a renewed interest in
quantitative aspects of these effects \cite{Bernreut,Schirkov,japrahy}.
There are two basic reasons for this, both of which are relevant to
the subject of this article.

The first concerns the exploitation of the complete NNLO QCD calculations
that have recently become available for
quantities such as \reff{Repem} \cite{Gor} or $\Gamma_h$
\cite{Lar1,Lar2} and which exist basically for
massless quarks only. The NNLO corrections are tiny effects and to
include them makes sense only if they are large compared to errors
resulting from the approximate treatment of the quark mass thresholds in
massless QCD. To quantify the importance of the NNLO correction to the
couplant $\alpha_s$, consider the difference between the values of
$\ase(\bbare{MS},M_Z)$ in the NLO and NNLO approximations,
assuming five massless quarks and taking, as an example,
$\Lambda^{(5)}_{\bbare{MS}}=0.2$ GeV. At the NLO the couplant $a\equiv
\ase/\pi$ is given as a solution to Equation \reff{NNLO} for $c_2=0$
and we get
$a^{\mr{NLO}}(\bbare{MS},M_Z)=0.03742$. At the NNLO and in \bbar{MS} RS,
where $c_2(\bbare{MS},n_f=5)=1.475$, we find
$a^{\mr{NNLO}}(\bbare{MS},M_Z)=0.03665$. The relative difference between
these two approximations,
\be
\frac{a^{\mr{NLO}}-a^{\mr{NNLO}}}{a^{\mr{NLO}}}\doteq 0.02,
\label{relcon}
\ee
thus amounts to about 2\%. The NNLO correction should therefore be
included only if the neglected effects can be expected to be
smaller than this number. However, as we shall see, quark mass threshold
effects can in some circumstances be of just this magnitude!\footnote{
In general, as the magnitude of higher order corrections to \as depends
on
the renormalization scheme employed, so does also the estimate
\reff{relcon}. However, if defined as the relative difference between
the NLO and NNLO approximations, this dependence is not strong
(see the concluding paragraph of Section 4).}

The second reason is related to the problem of comparing the values
of $\alpha_s$,
determined from different quantities characterized by vastly
different momentum scales. As recently emphasized in an extensive
review of \as determinations \cite{Bethke}, there is a small, but
nonnegligible discrepancy between the
value of $\ase(\bbare{MS},M_Z)$ obtained by extrapolation from some of
the low energy quantities, and  $\ase(\bbare{MS},M_Z)$ determined
directly at the scale $M_Z$ at LEP -- the latter giving the value higher
by about 5--10\%. Simultaneously, it has been noted in ref. \cite{Bethke}
that there is an exception to this behaviour in the case of the ratio
$R_{\tau}$, which, when extrapolated from $m_{\tau}$ to
$M_Z$, gives values of \as close to those measured directly at LEP
($0.120\pm0.005$ \cite{Pich2}). The physical relevance of the
discrepancy between the low energy extrapolations and direct
measurements of \as at LEP has very recently been emphasized
in Ref. \cite{Schifman}. In particular, its author argues that the
extrapolation of $\ase(m_{\tau})$ to $\ase(M_Z)$ is unreliable due to
limited control of the power corrections. The question of estimating
the theoretical uncertainty in the extraction of $\ase(m_{\tau})$ from
data on $R_{\tau}$ is also discussed in ref. \cite{Alta}
As shown in \cite{japrahy}, a part of this overestimate of the
extrapolated value of $\ase(M_Z)$ pointed out in \cite{Schifman} may
in fact be due to the approximate treatment of the $c$ and $b$ quark
thresholds.

We shall now analyze the quantitative consequences
of the exact treatment of quark mass thresholds at the LO and formulate
the conventional matching procedure \cite{Mar} for massless quarks in
such a way that its results are so close to
those which are exact that
the available NNLO calculation can be consistently included.
We shall describe in detail the approximation in which the $u,d$ and
$s$ quarks are considered massless while the $c,b$ and $t$ quarks
remain massive.

As complete
multiloop calculations with massive quarks are very complicated and
available only at the leading order, all higher order
phenomenological analyses use the calculations with a fixed effective
number $n_f$ of massless quarks, depending on the characteristic scale
of the quantity. For relating two regions of different effective
numbers of massless quarks the approximate matching procedure developed
in Ref. \cite{Mar}
is commonly used. It should be emphasized that this
procedure concerns only those mass effects that can be absorbed in the
renormalized couplant.  At higher orders there are, however, mass
effects that remain in the expansion coefficients even after the effects
of heavy quarks have been absorbed in a suitably defined running
couplant.

In QCD with massive quarks the renormalization group
equation for the couplant $a(\mu)$ formally looks the same
as in massless QCD.
The only, but important, difference concerns the two lowest order
$\beta$--function coefficients, $b,\beta_1=bc$, which are no longer
unique as in massless QCD, but may depend on the scale $\mu$.
While in the class of \bbar{MS}-like renormalization conventions
$b=11/2-n_f/3$, exactly
 as in massless QCD, in MOM-like ones it becomes a
nontrivial function of the scale $\mu$ \cite{RG}:
\be
b(\mu/m_i)=\frac{11}{2}-\frac{1}{3}\sum_{i} h_i(x_i),\;\;\;\;
x_i\equiv\frac{\mu}{m_i},
\label{xi}
\ee
where the sum runs over all the quarks considered, $m_i$ are the
corresponding renormalized quark masses\footnote
{For the purposes of this discussion quark masses can regarded
as constants. In numerical estimates
they are identified with the running masses at the scale $M_Z$.}
and the threshold function $h(x)$ is given as Ref. \cite{RG}.
The shape of the function $h(x)$ is actually not quite unique and
depends on the vertex chosen for the definition of the
renormalized couplant. This fact was first noted in Ref. \cite{Nacht}
and subsequently explained in Ref. \cite{Pol}. The  form of $h(x)$ used
below corresponds to quark-gluon-quark vertex with massless quarks,
using any momentum configuration and any invariant decomposition in
MOM-like RS \cite{Pol}. It is
appropriate for most of the extrapolations from low energy
quantities to LEP energy range. The same form is valid for the
ghost-gluon-ghost vertex. The three gluon vertex gives somewhat different
form of $h(x)$, though its behaviour for small and large $x$ is the
same. According to \cite{RG}:
\be
h(x)\equiv 6x^2 \int_0^1 \mr{d}z \frac{z^2(1-z)^2}{1+x^2 z(1-z)}=
1-\frac{6}{x^2}+\frac{12}{x^3\sqrt{4+x^2}}
\ln\frac{\sqrt{4+x^2}+x}{\sqrt{4+x^2}-x}
\doteq \frac{x^2}{5+x^2}.
\label{h(x)}
\ee
The last, approximate, equality is a very accurate
approximation of the exact form of $h(x)$ in the whole range
$x\in (0,\infty)$. This allows a simple treatment of the
quark mass thresholds at the LO. There is, unfortunately, no analogous
calculation of the next $\beta$--function coefficient, $bc$, for massive
quarks.  This is one of the reasons why most of the phenomenological
analyses use the so called `step' approximation, in which at any value
of $\mu$ one works with a finite effective number of massless quarks,
which changes discontinuously at some matching points $\mu_i$.
Consequently, $b(n_f)$ effectively becomes a function of $\mu$,
discontinuous at these matching points, as shown in Fig.1a.

 \begin{figure}
 \begin{center}
 \epsfig{file=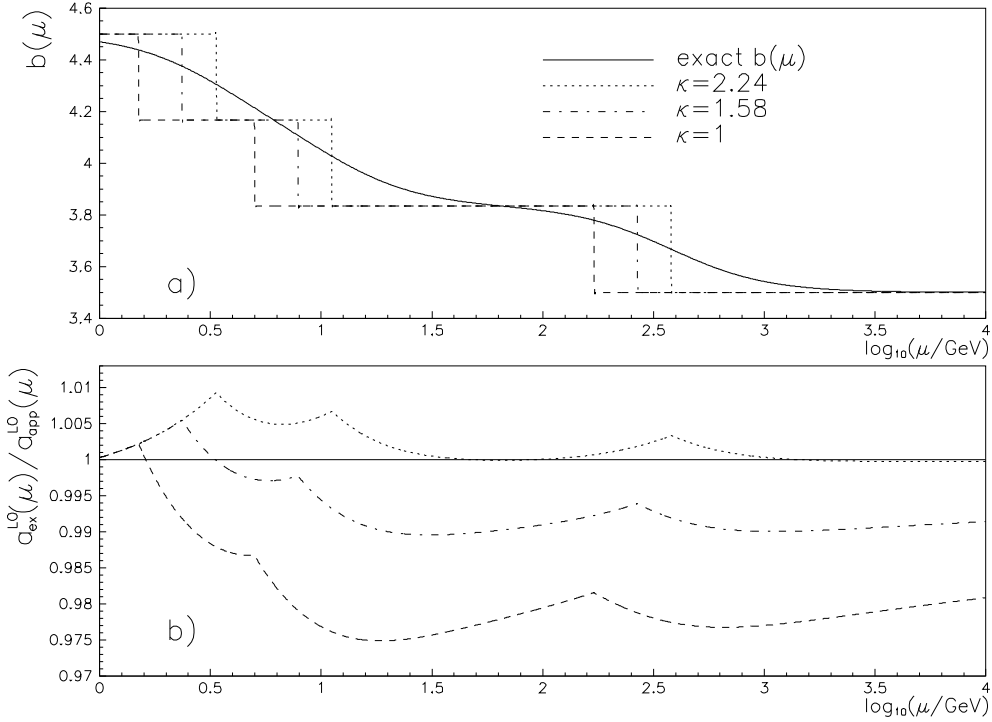,height=8cm}
 \end{center}
 \caption{a) $b(\mu/m_i)$ together with three step
 approximations, corresponding to matching at the points
 $\mu_{i}=\kappa m_i;i=c,b,t$ with $\kappa=1,1.58,2.24$;
 b) the ratio $R_a$ for the same three approximations, made to
 coincide at $\mu_0=1$ GeV.}
 \end{figure}

The matching points are assumed to be proportional to the masses of the
corresponding quarks, $\mu_i\equiv \kappa m_i$. In principle,
a different
quark threshold could be associated with a different $\kappa$, but for
simplicity's sake we take them equal. The
free parameter $\kappa$, allowing for the variation of the
proportionality factor, turns out to be quite important for the accuracy
of the step approximation. At the LO the matching procedure then
consists of
the following relations at the matching points $\mu_i$ (the numbers
in the superscript define the corresponding effective number of
massless quarks)\footnote
{The parameter $\Lambda$ appearing in this as well as the other
formulae in this section is the \underline{leading order} $\Lambda$
parameter, $\Lambda^{\mr{LO}}$, which cannot be associated with any
well-defined RS. As there is no danger of confusion, we shall
drop the superscript `LO' for the remainder of this section.}:
\begin{eqnarray}
a^{\mr{LO},3}_{app}(\kappa m_{c}/\Lambda^{(3)}) & = &
a^{\mr{LO},4}_{app}(\kappa m_{c}/\Lambda^{(4)})
\Rightarrow \Lambda^{(4)}=\Lambda^{(3)}\left(\frac{\Lambda^{(3)}}
{\kappa m_c}\right)^{1/3b(4)}
\label{mc} \\
a^{\mr{LO},4}_{app}(\kappa m_{b}/\Lambda^{(4)}) & = &
a^{\mr{LO},5}_{app}(\kappa m_{b}/\Lambda^{(5)})
\Rightarrow \Lambda^{(5)}=\Lambda^{(4)}\left(\frac{\Lambda^{(4)}}
{\kappa m_b}\right)^{1/3b(5)}
\label{mb} \\
a^{\mr{LO},5}_{app}(\kappa m_{t}/\Lambda^{(5)}) & = &
a^{\mr{LO},6}_{app}(\kappa m_{t}/\Lambda^{(6)})
\Rightarrow \Lambda^{(6)}=\Lambda^{(5)}\left(\frac{\Lambda^{(5)}}
{\kappa m_t}\right)^{1/3b(6)}.
\label{mt}
\end{eqnarray}
Note that each of the intervals of fixed $n_f$
is associated with a different value of the
$\Lambda$--parameter, $\Lambda^{(n_f)}$. The resulting dependence
of $a(\mu/m_i)$ on $\mu$ is thus continuous at each of the
matching points, but its derivatives at these points are
 discontinuous, reflecting the discontinuity of the step
 approximations to $b(\mu/m_i)$.
This procedure can easily be extended to any finite order. Let us
point out that the more sophisticated procedure for matching the
couplants corresponding to different effective $n_f$ developed in
Ref. \cite{BW} coincides in the LO with the above relations (48--50).
To estimate the errors in \as resulting from the
above-defined approximate treatment of quark thresholds we merely need
solve the LO equation with exact explicit mass
dependence as given in \reff{xi}:
\be
\frac{\mr{d}a(\mu)}{\mr{d}\ln \mu}=-a^2\left(\frac{11}{2}-\frac{1}{3}
\sum_{i=1}^{6} h(x_i)\right).
\label{RGhx}
\ee
For our purposes the approximation $h(x)\doteq x^2/(5+x^2)$
is entirely adequate and yields
\be
a(\mu)=\frac{1}{\ds {\left(\frac{11}{2}-\frac{3}{3}\right) \ln
\frac{\mu}{\Lambda^{(3)}} -\frac{1}{3} \sum_{i=c,b,t}\ln
\frac{\sqrt{\mu^2+5m_i^2}}
{\sqrt{\left(\Lambda^{(3)}\right)^2+5m_i^2}}}},
\label{am}
\ee
where the fraction $\frac{3}{3}$ comes from the sum over the three
massless quarks $u,d$ and $s$ and $\Lambda^{(3)}$ is the corresponding
$\Lambda$--parameter appropriate to three massless quarks.
For the heavy quarks $c,b$ and $t$ we take in the following $m_c=1.5$
GeV, $m_b=5$ GeV, $m_t=170$ GeV.  The distinction between the `light'
and `heavy' quarks is given by the relative magnitude of
$m_i$ and $\Lambda$, the latter being defined by the condition
$5m_i^2\gg\Lambda$. For the above values of $m_c,m_b$ and $m_t$
this condition
 is very well satisfied.  Consequently, for $\mu \ll m_i,i=c,b,t$
\reff{am} approaches smoothly $a^{\mr{LO}}$ for $n_f$=3, while for $\mu
\gg m_i$, and neglecting $\Lambda^{(3)}$ with respect to $5m_i^2$,
it goes to
\be
a(\mu)=\frac{1}{\ds b(6)\ln \frac{\mu}{\Lambda^{(3)}}+ \frac{1}{3}
\ln\left( \frac{\sqrt{5}m_c}{\Lambda^{(3)}}
\frac{\sqrt{5}m_b}{\Lambda^{(3)}} \frac{\sqrt{5}m_t}{\Lambda^{(3)}}
\right)}= \frac{1}{\ds b(6)\ln \frac{\mu}{\Lambda^{(6)}
(\sqrt{5})}},
\label{acbt}
\ee
where the parameter $\Lambda^{(6)}(\kappa)$ depends in general on
$\kappa$ and
\be
\Lambda^{(6)}(\sqrt{5})
\equiv \Lambda^{(3)}
\left(\frac{\Lambda^{(3)}}{\sqrt{5}m_c}
\frac{\Lambda^{(3)}}{\sqrt{5}m_b}
\frac{\Lambda^{(3)}}{\sqrt{5}m_t}\right)
^{\frac{1}{3b(6)}}
=\left(\frac{1}{\sqrt{5}}\right)^{\frac{1}{b(6)}}
\Lambda^{(6)}(1)
\label{Lambda6}
\ee
coincides with $\Lambda^{(6)}$ defined via the
subsequent application of the matching
relations
(48--50) for $\kappa=\sqrt{5}\doteq 2.24$.
Even though from the point of view of the matching
procedure, $\kappa$ is not exactly fixed, the value $\kappa=\sqrt{5}$
will be shown to be in a certain sense the best choice.
The relation between $\Lambda^{(6)}$ and $\Lambda^{(3)}$
depends nontrivially on $\kappa$.

In Fig. 1a, $b(\mu)$ is plotted as a function of $\mu$ for the
above-mentioned
masses of $c,b$ and $t$ quarks, together with its step
approximations and corresponding to three different values of
$\kappa=1,\sqrt{5},\sqrt{5/2}$.
There is hardly any sign of the steplike behaviour of the function
$h(x)$ in the region of the $c$ and $b$ quark thresholds and only a
very unpronounced indication of the plateau between the $b$ and $t$
quark thresholds. The step approximations are poor representations
of the exact $h(x)$,
primarily due to the rather slow approach of $h(x)$
to unity as $x\rightarrow \infty$. However, there is a marked
difference between the three approximations. While the step
approximation with the conventional choice $\kappa=1$
understimates the true $h(x)$ in the whole interval displayed, and would
do so even when some smoothing were applied, $\kappa=\sqrt{5}$ gives
clearly much better approximation as the corresponding curve is
intersected by the exact $h(x)$  at about the middle of each step.

In Fig. 1b the $\mu$ dependence of the ratio
\be
R_a\equiv\frac{a^{\mr{LO}}_{ex}(\mu)}{a^{\mr{LO}}_{app}(\mu)},
\label{app}
\ee
between the above exact solution \reff{acbt} and the approximate
expressions for the above mentioned values of $\kappa$,
is plotted assuming $\Lambda^{(3)}=200$ MeV\footnote{
The resulting ratio $R_a(M_Z)$ depends only weakly on
$\Lambda^{(3)}$. Note, however, that the current analysis of the
LEP data gives the value of $\Lambda^{(3)}$ in the range 500--700 MeV
in disagreement with the QCD sum rules analysis \cite{Schifman}.}.
 As we basically want to compare the results of
different extrapolations starting from the same initial $\mu_0$,
$\Lambda^{(3)}$ used in the approximate solutions
has been rescaled by the
factor 1.004 with respect to $\Lambda^{(3)}$ in \reff{am},
thereby guaranteeing that all expressions coincide at $\mu_0=1$ GeV.
Any deviation from unity in Fig. 1b is then entirely the effect
of an approximate treatment of the heavy quark thresholds. Figure 1
contains several simple messages.

The approximate solutions based on the matching procedure
defined in (48--50)
are generally much
better immediately {\em below} the matching point than
above it, and worst at about $5m_{\mr{match}}$. This reflects the
fact that the function $h(x)$ vanishes fast (like $x^2$) at zero
but approaches unity for $x\rightarrow \infty$
only very slowly. Moreover, in the $M_{Z}$ range
the effect of the $c$ quark threshold is essentially the same as
that of the $b$ quark and both are much more important than that of the
top quark, although $M_Z/m_c\approx 60$, $M_{Z}/m_b\approx 18$, while
$M_{Z}/m_t \approx 1/2$!

The effect of varying $\kappa$ is quite important, in particular
with respect to the $c$ and $b$ quark thresholds.
In general, $\kappa >1$ improves the
approximation above,
but worsens it immediately below  the matching point
it. The choice $\kappa=\sqrt{5}$, suggested by the asymptotic behaviour
of \reff{acbt}, is clearly superior
in practically the whole displayed interval $\mu\in(1,10^4)$ GeV
and leads to an excellent (on the level of 0.1\%) agreement
with the exact solution in this interval.
On the contrary, the conventional choice $\kappa=1$
leads to a much larger deviation from the exact result, which exceeds
$2$\% in most of this region. This discrepancy is of the same magnitude
as the effects of NNLO corrections to the couplant itself.
It thus turns out that the effect of an exact treatment of the quark
mass thresholds for the \underline{extrapolation} of \as from the scales
around $\mu\approx 2$ GeV up to $\mu=M_Z$ is as important as that of the
NNLO correction to \as and must therefore be taken into account
whenever the latter is considered and compared with
\as determined at these vastly different scales.

 On the other hand, the preceding discussion tells us little about
the accuracy of the approximation of five massless quarks directly
\underline{at} the scale $M_Z$, for instance when calculating
$\Gamma_h$. This question will be addressed in section 5.

The analysis of quark mass effects in \as presented above
strictly speaking holds only for the LO. Nevertheless, as both the
mass effects and the higher order perturbative corrections are small
effects, it seems reasonable to expect that the conclusions drawn
in this section will have more general validity.

\section{Application to $\Gamma_h$}
We now come to the quantitative estimate of the
theoretical uncertainties of perturbative QCD predictions
for $\Gamma_h$. In the preceding section we
discussed the approximation in which \as is given by an expression
corresponding to five massless quarks. In this section we quantify
the uncertainties resulting from the RS ambiguity, discussed in Section
2. The nontrivial dependence of $\Gamma_h$ on $m_b$ and
$m_t$, coming from effects not included in the running couplant
corresponding to five massless quarks, is
discussed in the next section. Nevertheless in order to avoid
unnecessary repetition we include the dependence on the ratio
$M_Z/m_t$ already in the formulae quoted below and obtained
recently in Refs.
\cite{Gor,Lar1,Lar2,CKT,Kn}. The dependence of the expansion
coefficients on the bottom quark mass
has extensively been studied in Ref. \cite{ChKK}. We do not
write it out here explicitly, as this would further complicate
the structure of \reff{Gamma}, each of the coefficients
in \reff{Gamma} becoming a different function of the ratio $m_b/M_Z$.
At the end of this section we
shall merely recall the leading contributions to $R_Z$ coming
from $m_b/M_Z$ terms and discuss their numerical importance.

In order to quantify the theoretical uncertainty related to the choice
of the RS, we first define the set of `allowed' RS. As
emphasized above, this is to large extent a subjective matter. Based on
our previous experience we {\em define} as a measure of this
uncertainty the difference between the results obtained (for the same
$\Lambda_{\overline{\mr{MS}}}$) in the three principal methods
set out in Section 3:
PMS, ECH and $\overline{\mr{MS}}$. This choice is to large extent
arbitrary, but as the $\overline{\mr{MS}}$ RS is used in
most phenomenological analyses, we adopt it for the lack of anything
better. The formulae quoted below are taken from Ref. \cite{Lar2}.

The basic quantity of interest, $R_Z$, defined in \reff{R},
has a nontrivial structure which mixes the effects of electroweak
interactions with those of pure QCD. It can be written
in the following decoupled form (i.e., for five massless flavours and
the explicit $m_t$ dependence of the expansion coefficients)
as the sum of three terms with different electroweak factors
\cite{Heb} and separated further into four possible combinations of
vector, axial vector and singlet, nonsinglet contributions:
\begin{eqnarray}
R_Z & = & \left(R^{V,NS}+R^{A,NS}\right)
  +R^{V,S}+R^{A,S}   \nonumber\\
  & = & \sum^{5}_{i=1}\left(g^2_{V,i}+g^2_{A,i}\right)
\left[1+a^{(5)}+(a^{(5)})^2 r_1+(a^{(5)})^3 r_2 \right] \nonumber \\
 &  & + \left(\sum_{i=1}^5 g_{V,i}\right)^2\left[(a^{(5)})^3 s_3+
        (a^{(5)})^3 s_3^{top}
\right] \nonumber \\
   &  & + \left(\frac{1}{4}\right)
\left[(a^{(5)})^2 t_2+(a^{(5)})^3 t_3\right],
\label{Gamma}
\end{eqnarray}
where  $g_{V,i}=t_{3,i}-2Q_{i}\sin^2\theta_{W},\;g_{A,i}=t_{3,i}$,
$t_3$ is the third component of the weak isospin and
the sums over the electroweak coupling constants equal
\be
\Gamma_1 \equiv \sum^{5}_{i=1}\left(g^2_{V,i}+g^2_{A,i}\right)=
\frac{5}{2}+\frac{44}{9}\sin^4\theta_{W}-\frac{14}{3}\sin^2\theta_W
=1.6807\pm0.0012
\label{Gamma1}
\ee
\be
\Gamma_2 \equiv
 \left(\sum_{i=1}^5 g_{V,i}\right)^2=\left(\frac{1}{2}+\frac{2}{3}
\sin^2\theta_W\right)^2=0.42850\pm0.00028.
\label{Gamma2}
\ee
The last equalities in the above relations correspond to the
world average $\sin^2\theta_W=0.2329\pm0.0005$. Note that \reff{Gamma1}
implies
\be
 \frac{\Delta\Gamma_2}{\Gamma_2}\doteq 6.5\;10^{-4}<
 \frac{\Delta\Gamma_1}{\Gamma_1}\doteq 7.1\;10^{-4}\ll
 \frac{\Delta \Gamma_h}{\Gamma_h},
\label{chyby}
\ee
which, combined with  \reff{errors}, means that the theoretical
uncertainties of QCD predictions -- in the square brackets in
\reff{Gamma} -- should be compared to the
error of $R_Z$ itself.

The expansion coefficients entering the above formula can be expressed
as functions of the ratio $x\equiv M_{z}/m_t(M_Z)$, where $m_t(M_Z)$ is
the renormalized, `runnning' mass of the top quark, taken at the scale
$M_Z$. For $n_f=5$ and in small $x$ expansions we have \cite{Lar2}:
\begin{eqnarray}
r_1 & = & 1.409  \nonumber \\
   & & +[0.065185-0.014815\ln x]x  \nonumber \\
   & & +[-0.0012311+0.00039683\ln x]x^2 \nonumber \\
  & & +[0.000061327-0.000023516\ln x]x^3 +{\cal O}(x^4) \\
r_2 & = & -12.767 \nonumber  \\
  & & +[-0.17374+0.21242\ln x -0.037243\ln^2 x]x \nonumber \\
  & & +[-0.0075218-0.00058859\ln x+0.00038305\ln^2 x]x^2 \nonumber \\
  & & +[0.00050411-0.00012099\ln x+0.000031419\ln^2 x]x^3+{\cal O}(x^4)
  \\
s_3  & = & -0.41318 \nonumber \\
s_3 ^{top} & = & 0.027033x+0.036355x^2+0.00058874x^3+{\cal O}(x^4)
\\
t_2 & = & -3.0833+\ln x+0.086420x+0.0058333x^2+0.00062887x^3+
{\cal O}(x^4) \\
t_3 & = & 18.654+1.7222\ln x+1.9167\ln^2 x \nonumber  \\
  & & +[-0.12585+0.28646\ln x-0.011111\ln^2 x]x \nonumber \\
  & & +[-0.0031322+0.012117\ln x-0.0011905\ln^2 x]x^2 \nonumber \\
  & & +[-0.00088827+0.00047262\ln x-0.00017637\ln^2 x]x^3+{\cal O}(x^4).
\end{eqnarray}
Because of different electroweak factors in front of them, each of the
expressions in the square brackets of \reff{Gamma} is separately from
the point of view of QCD RS invariant.
As the optimization according to either the PMS or ECH
methods does not commute with the operation of addition, the first
question we have to answer is the order of these operations.
In the absence of uncertainties in the values of the
electroweak factors, the proper way would be first to sum all three
terms in \reff{Gamma} and then to fix the RS. In reality, however, the
errors of electroweak factors induce uncertainties in values of the
coefficients multiplying powers of the QCD couplant. To optimize, in one
way or another, the resulting QCD perturbative expansion \reff{Gamma} in
such circumstances is not a well-defined exercise and we therefore have
chosen to follow the opposite route. We believe that to get an estimate
of the RS dependence of QCD calculations this second route is adequate.
 \begin{figure}
 \begin{center}
 \epsfig{file=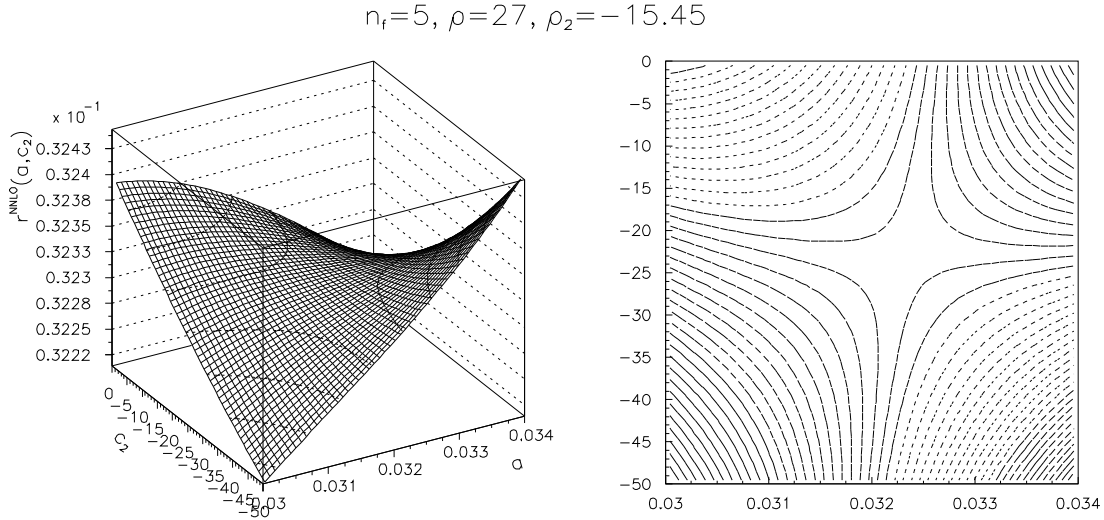,width=15cm}
 \end{center}
 \caption{The NNLO approximant $r^{(3)}(a,c_2)$ in the vicinity of
 the saddle point for $\rho=23$.}
 \end{figure}
In this section we shall discuss the numerical importance of the first
two terms in \reff{Gamma} for $m_t=0$. The effects of non-zero $m_t$,
as well as the third term in \reff{Gamma}, the existence of which is
also closely related to non-zero $m_t$, are dealt with in the next
section.
The second term of \reff{Gamma}, multiplied by \reff{Gamma2}, is given
at the LO only and no optimization is possible. Fortunately it
contributes, in $\overline{\mr{MS}}$ RS, a mere $10^{-5}$ to $R_Z$
and is thus clearly negligible.

The dominant contribution to $R_Z$ comes from the first term in
\reff{Gamma}. The term in the square brackets can be written as
$1+r(M_Z)$, where $r(Q)$ has exactly the form of \reff{r(Q)} and,
moreover, for massless quarks coincides with the above expression for
\reff{Repem} \cite{Gor}.  For $n_f=5$ the crucial RG invariant
$\rho_2=-15.45$, which implies (for detailed discussion see
Ref. \cite{my}) that the saddle
point of the NNLO approximant $r^{(3)}(\rho)$ will occur at the point
$(a_{\mr{PMS}},c_2^{\mr{PMS}})$, where $c_2^{\mr{PMS}}(\rho)\rightarrow
1.5\rho_2\doteq -23.2$. In Fig. 2 $r^{(3)}(\rho=27)$ is plotted as a
function of $a,c_2$ near this saddle point, together with the contours
of the constant $r^{(3)}$.
We have calculated the NLO as well as the NNLO approximants of the
quantity in the first square bracket of \reff{Gamma}, with the unity
subtracted, in the three chosen RSs
and in the interval $\rho\in(18,28)$,
which corresponds to the measured value of $M_Z\doteq 91.4$ GeV and
$\Lambda_{\overline{\mr{MS}}}^{(5)}$ in the interval
$\Lambda_{\overline{\mr{MS}}}^{(5)} \in(50,500)$ MeV. As the differences
are tiny we normalize all our results to the NLO result in the
conventional $\overline{\mr{MS}}$ RS, and plot the relative difference,
\be
r_{\rho}\equiv
\frac{r^{(i)}(\mr{RS})}{r^{\mr{NLO}}(\Lambda_{\overline{\mr{MS}}})}-1,
\label{rat}
\ee
where i=NLO, NNLO and RS=PMS, ECH or $\overline{\mr{MS}}$.
We draw the following conclusion from Fig. 3:
\ben
\item
The differences between PMS and ECH approaches are minuscule, about
0.1\% at the NLO and totaly negligible at the NNLO,
\item the difference between the PMS (or ECH) and $\overline{\mr{MS}}$
  approaches is
   \bit
     \item about 0.7\% at the NLO and
     \item about 0.3\% at the NNLO (this comes from the ratio of
     the dotted and dash--dotted curves in Fig. 3). This documents the
     trend, observed in earlier works, that inclusion of the NNLO
     corrections \underline{diminishes} the RS dependences and thus
     decreases the theoretical uncertainty.
   \eit
\item the differences between the NLO and NNLO approximations
amount to about 2\% in the \bbar{MS} RS (as already estimated in
Section 3) and to about 3\% for PMS (or ECH).
\een
  \begin{figure}
 \begin{center}
 \epsfig{file=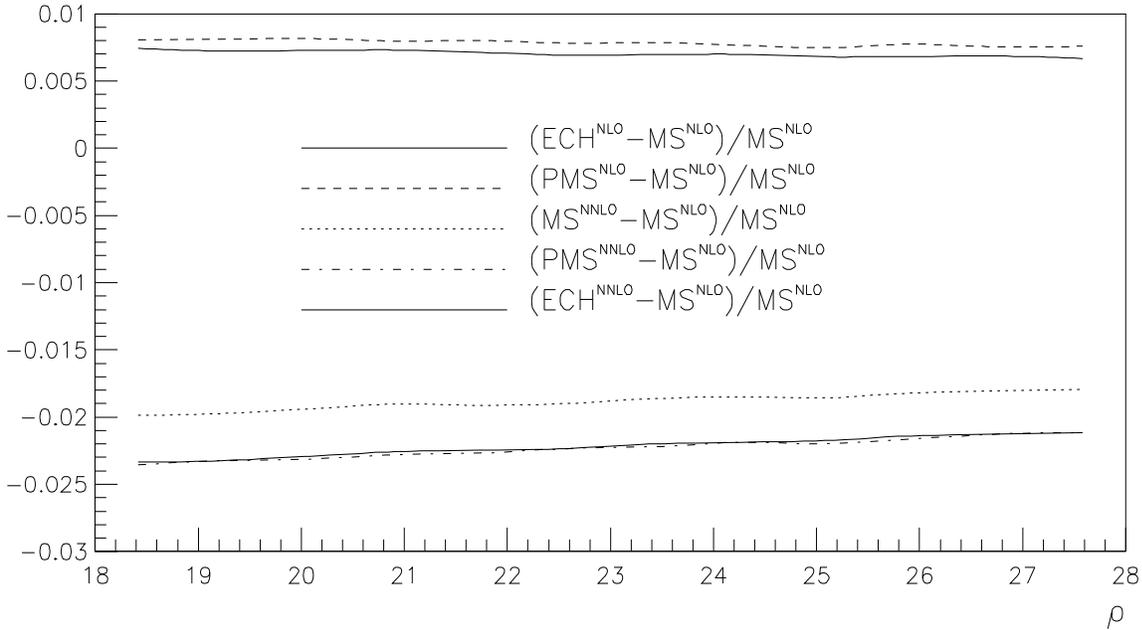,height=6.5cm,width=15cm}
 \end{center}
 \caption{The ratio $r_{\rho}$, defined in the text.}
 \end{figure}
Given
the current precision of the data \reff{errors}, which translates into
8.5\% accuracy on the couplant $a(M_Z)$, this implies that none of the
mentioned differences is discernible.  Moreover, only the last
difference, namely the effect of including the NNLO correction,
has any chance of being seen in the data, but even this would require
at least a factor three improvement in the precision of $\Gamma_h$.

\section{The dependence of $R_Z$
on $m_b$ and $m_t$}
As already pointed out, massive quarks complicate the
consistency conditions, discussed in Section 2, as their
renormalized masses also `run' and there is so far no generalization
of the relations \reff{r1r2} available. We therefore estimate the
effects of non-zero top and bottom quark masses within the
$\overline{\mr{MS}}$ RS. Recall that \reff{Gamma}
are expansions in $a^{(5)}$ corresponding to five massless quarks,
 and thus all effects of finite $m_b$
and $m_t$ are contained in the expansion coeffiecients.

First we deal with the effects of the finite $x=M_Z/m_t$ in the first
and third term of \reff{Gamma}.
In the first case the $x$-dependent terms of $r_1$ and $r_2$ vanish
in the limit $x\rightarrow 0$ and the leading, constant, ones
correspond to $n_f=5$ in accordance with the decoupling theorem
\cite{AC}. For $x=0.4225$, corresponding to $M_Z=91.187$ and
$m_t(M_Z)=140$ GeV, as suggested in Ref. \cite{Lar2},
$R^{V,NS}+R^{A,NS}$ changes by the factor
$3.7\;10^{-5}$, which is an effect two orders of magnitude smaller than
the experimental error of $R_Z$, estimated in \reff{errors}, and one
order of magnitude smaller than the effects discussed in the preceding
section. It can therefore be safely neglected.

The contribution of the last term in \reff{Gamma}, coming from the
axial vector, flavour singlet channel, where the familiar axial anomaly
operates, is numerically nonnegligible. Despite the fact
that the leading terms in $t_2$ and $t_3$ are $x$-independent, the
nonvanishing of $R^{A,S}$ is  due to non-zero
$m_t$, or, more precisely, to
the effect of the difference $m_{t}-m_{b}$
\footnote{An alternative interpretation of this effect is discussed in
Ref. \cite{JaHor}.}.
 This comes from the fact
that axial couplings of quarks in weak dublets are opposite and,
apart from mass effects, their
contributions cancel. Consequently,
for $m_t=0$ both $t_2$ and $t_3$, and thereby also $R^{A,S}$, must
vanish.  The fact that this vanishing is not obvious in the small $x$
expansions of \reff{Gamma} is not surprising, as $m_t\rightarrow 0$
corresponds on the contrary to $x\rightarrow \infty$.
Note that the presence of powers of $\ln x$ in the coefficients $t_2$
and $t_3$ implies that a resummation of these logarithms is necessary
\cite{Kn} even in the limit $x\rightarrow 0$.
 As the contribution $R^{A,S}$ can be interpreted as the top quark mass
effect, we again determine its contribution in the $\overline{\mr{MS}}$
RS only. For $m_t(M_Z)=140$ GeV, $R^{A,S}=-0.00160$.

The effects of finite value of $m_b/M_Z$ have been studied in detail
in \cite{ChKK}. Here we merely recall the form and numerical values
 of the
leading contributions in the vector and axial vector, nonsinglet
$b\bare{b}$ channels resulting from these effects:
\begin{eqnarray}
R^{V,NS}_{b} & \equiv & g^2_{V,b}\left(1+\left[
1+12\frac{m_b^2(M_Z)}{M_Z^2}\right]a^{(5)}(M_Z)+
{\cal O}([a^{(5)}]^2)\right),
 \label{RVb} \\
R^{A,NS}_{b} & \equiv & g^2_{A,b}\left(1-6\frac{m^2_b(M_Z)}{M_Z^2}
+{\cal O}(a^{(5)})\right).
\label{RAb}
\end{eqnarray}
For $m_b(M_Z)=4.8$ \cite{ChKK} and $a^{(5)}(M_Z)=0.037$ the
leading $m_b/M_Z$ contribution to $R_Z$ coming from the axial channel
thus amounts to $4.2~10^{-3}$, whereas
in the vector channel we get a mere $1.5~10^{-4}$.

\section{The estimate of the still higher order terms}
As we saw in the preceding two sections, the NNLO corrections to
$\Gamma_h$ are, compared to experimental
accuracy of its measurement, tiny effects, and so it seems reasonably
safe to stop at this order. On the other hand
it is generally accepted that perturbative series in QCD do not converge,
but represent  merely asymptotic
expansions to the full result. In such situation it is certainly useful
to have at least some estimate of the magnitude of the
so far uncalculated (and in the near future uncalculable) higher orders.

An attempt in this direction has recently been made in \cite{KatSt},
using the so-called `improvement formula' of Ref. \cite{PMS}, which
represents an approximation of the PMS optimization discussed in
subsection 2.3. Its essence is to reexpand the PMS result optimized to
the Nth order in powers of the couplant in any fixed RS\footnote
{The optimized result is, of course, independent of the
choice of this RS and is constructed from quantities up to the Nth
order only.}
and take the coefficient of this expansion at the (N+1)th
order as an estimate of its true value. Instead of the PMS approach
the ECH one of subsection 2.4 can be equally well used for this
purpose. The resulting estimates are only slightly different.

Here we outline the main steps of this method for the quantity
\reff{Repem}, closely related to $\Gamma_h$.
Consider first the Nth order partial sum of the perturbative expansion
for a physical quantity $D$\footnote
{In this section we drop the specification of the number of
quark flavours in the couplant $a$.}:
\be
D_{N} = \sum^{N-1}_{i=0} \, d_{i} a^{i+1}.
\label{1}
\ee
Carrying out the optimization of \reff{1} according to either
the PMS or ECH
approaches leads to the optimized result, denoted below as
$D_N^{\mr{opt}}(a_{\mr{opt}})$. If we now reexpand $D_{N}^{\mr{opt}}
(a_{\mr{opt}})$ in terms of the couplant $a(\mr{RS})$ in a chosen RS
\footnote{The magnitude and therefore also the estimate of higher order
coefficients depends on the choice of RS.
We drop the argument `RS' of $a(\mr{RS})$ here.} we find
\be
D_{N}^{\mr{opt}} (a_{\mr{opt}}) = D_{N} (a) + \delta D_{N}^{\mr{opt}}
a^{N+1},
\label{3}
\ee
where
\beq
\delta D_{N}^{\mr{opt}} = \Omega_{N}(d_{i}, c_{i}) -
\Omega_{N} (d_{i}^{\mr{opt}}, c_{i}^{\mr{opt}})
\label{4}
\eeq
give, according to ref.
\cite{KatSt}, the estimate of the coefficient $d_{N}$
in the chosen RS. For the three lowest orders the functions
$\Omega_{N}(d_i,c_i)$ are given as \cite{KatSt}:
\begin{eqnarray}
\Omega_{2}& =& d_{0}d_{1} (c_{1} + d_{1}) \nonumber \\
\Omega_{3}& =& d_{0}d_{1} (c_{2} - \frac{1}{2} c_{1}d_{1}
-2d_{1}^{2} + 3d_{2}) \label{new1} \\
\Omega_{4}&=&\frac{d_0}{3} ( 3c_{3}d_1+c_2d_2-4c_2d_1^2+2c_1d_1d_2
-c_1d_3 \nonumber \\
& &+14d_1^4-28d_1^2d_2+5d_2^2+12d_1d_3 ).
\nonumber
\end{eqnarray}
These formulae can be derived from the following exact equations
relating $\Omega_j$ to the coefficients $d_j,c_j$ and the RG invariants
$\rho_j$:
\be
d_j=
\frac{\rho_j}{j-1}-\frac{c_j}{j-1}+\frac{\Omega_j}{d_0}.
\label{exact}
\ee
Note that in order to evaluate $\Omega_{j}$ only $d_i,c_i$ at
\underline{lower} orders $i\le j-1$ are actually needed!

If the ECH approach is used for the optimization, the formulae \reff{4}
is particularly simple as $d_i^{\mr{opt}}=0$ by definition and thus
\begin{eqnarray}
\delta D_{2}^{\mr{ECH}} &=& \Omega_{2} (d_{1}, c_{1}) \nonumber \\
\delta D_{3}^{\mr{ECH}} &=&
\Omega_{3}(d_{1}, d_{2}, c_{1}, c_{2}) \label{new2} \\
\delta D_{4}^{\mr{ECH}}&=&\Omega_4(d_1,d_2,d_3,c_1,c_2,c_3).
\nonumber
\end{eqnarray}
The estimate \reff{new2} is thus equivalent to the assumption
that $d_N$ is dominated by the last term in \reff{exact}. Extensive
discussion of this assumption and its consequences is given in
Ref. \cite{KatSt} and is also related to Ref. \cite{beta}.

Using the PMS approach the resulting estimate of $d_N$ differs
from \reff{new2} by the presence of the second term in \reff{4}, which
does not vanish as in the ECH approach. However, it was shown in
ref. \cite{KatSt} that $\Omega_2(d_i^{\mr{PMS}},c_i^{\mr{PMS}})$ and
$\Omega_4(d_i^{\mr{PMS}},c_i^{\mr{PMS}})$ are small and
$\Omega_3(d_i^{\mr{PMS}},c_i^{\mr{PMS}})=0$. In the following numerical
estimates only the ECH-based results are therefore presented.
 \begin{table}
\begin{center}
\begin{tabular}{|c|c|c|c|c|} \hline
$n_f$ & $r^{exact}_{2}$ & $r^{estimate}_{2}$ &
$r^{est}_{3}$ & $r_4^{estimate}-c_3r_1$\\ \hline\hline
1 & -7.84 & -14.41 &  -166 & -1750\\ \hline
2 & -9.04 & -12.65 &  -147 & -1161 \\ \hline
3 & -10.27 & -11.04 &  -128 & -668 \\ \hline
4 & -11.52 & -9.59 &  -112 & -263 \\ \hline
5 & -12.76 & -8.32 &  -97 & 67 \\ \hline
6 & -14.01 & -7.19 &  -83 & 330 \\ \hline
\end{tabular}
\end{center}
\caption{The estimate of the so far uncalculated higher order
coefficients for the quantity $R(s)$ in the \bbar{MS} RS and using the
ECH optimization procedure.}
\end{table}

There is one subtle point in the derivation of estimates for higher
order coefficients $r_k$ of time-like quantities like \reff{R} or
\reff{Repem}. For instance, $R(s)$ of \reff{Repem} is related to the
so-called $D$-function $D(Q^2)$, defined primarily in the Euclidean
region, via the dispersion relation
\be
D(Q^{2})=Q^{2}\int^{\infty}_{0}\,\frac{R(s)}{(s+Q^{2})^{2}}\mr{d}s.
\label{12}
\ee
The knowledge of $R(s)$ is in principle equivalent to that of $D(Q^2)$,
but as most of the optimization procedures or methods of higher order
estimates do not commute with the functional on the r.h.s. of \reff{12},
we face the question as to which quantity to apply Eq.(71).
This nontrivial problem is discussed in Ref. \cite{KatSt}, the
conclusion being that
they should be applied to the quantities in the Euclidean region --,
for instance, $D(Q^2)$. Having obtained the estimates for higher
order coefficients $d_j$, the corresponding estimates for the
coefficients $r_j$ of $R(s)$ follow from the relations
\bea
r_{1} & = & d_{1} \nonumber \\
r_{2} & = & d_{2} - \frac{\pi^{2} b^2}{12}, \nonumber \\
r_{3} & = & d_{3} - \frac{\pi^{2} b^2}{4} \nonumber \\
r_{4} &=& d_{4} -\frac{\pi^2b^2}{4} (2d_2+\frac{7}{3}c_1d_1
+\frac{1}{2}c_1^2+c_2)+\frac{\pi^4b^4}{80}.
\label{r4}
\eea
The terms, proportional to powers of $\pi^2$, come from the analytical
continuation of the couplant $a(\mu)$ from the Euclidean region, where
$\mu^2<0$ to the Minkowskean one, where $\mu^2>0$.
Taking into account the fact that in the \bbar{MS} RS we have
\cite{bbi}:
\begin{eqnarray}
d_{1}(\overline{\mr{MS}}) & \approx & 1.986-0.115 n_f \nonumber \\
d_{2}(\overline{\mr{MS}}) & \approx & 18.244-4.216 n_f+0.086 n_f^{2}
\label{14} \\
c_{2}(\overline{\mr{MS}}) & = & \frac{77139 - 15099 n_f + 325 n_f^{2}}
{9504 - 576 n_f},
\end{eqnarray}
and using \reff{new2} we get the estimates, obtained originally in Ref.
\cite{KatSt}, summarized in Table 2 and valid for the
\bbar{MS} RS.\footnote
{Neglecting the terms of the light-by-light type, which
violate the structure of \reff{Repem}.}
In order to get some feeling of the possible accuracy of these
estimates the above table also includes the results for the NNLO
coefficient $r_2$, for which the exact calculations are available.
In the case of the coefficient $r_4$, only the estimate for
the combination $r_4-r_1c_3$ is presented, as the four loop
$\beta$--function coefficient $c_3(\bbare{MS})$ is so far unknown.

As the dominant contribution to \reff{R} comes from the nonsinglet
channel -- first term of \reff{Gamma} --, the above estimates are
relevant
for this quantity as well. For $a(\bbare{MS},M_Z)=0.037$, $n_f=5$ and
using the estimates of Table 1, we find that the terms $r_3a^4$ and
$r_4a^5$ contribute approximately $-3~10^{-4}$ and $8~10^{-6}$
respectively\footnote
{For the latter contribution the additional assumption, concerning
$c_3(\bbare{MS})$, had been made: $c_3=c_2^2/c$.}.
Note that while the latter contribution is entirely
negligible, the former is of the same order as the RS uncertainty of the
NNLO contribution.

\section{Summary and conclusions}
In the preceding sections we have analyzed various contributions to, and
theoretical uncertainties of, the quantity \reff{R}. The results of
these analyses are summarized in Table 2.
All these numbers should
be contrasted with the current experimental error of $R_Z$, which is
$5.9~10^{-3}$.
We see that there is a number of affects that are comparable to the
current experimental accuracy of $R_Z$, the most important of them being
the NLO perturbative correction and, interestingly, the effects of finite
$b$ quark mass correction to the Born term in the axial channel. This,
however, has nothing to do with QCD.
On the other the hand, the data are not yet sufficiently precise to be
sensitive to, for instance, the NNLO perturbative correction.
Further improvement in the
measurement of $\Gamma_h$ is clearly very desirable.

\begin{table}
\begin{center}
\begin{tabular}{|l|l|l|} \hline
 & type of contribution &  contributes to $R_Z$ \\ \hline\hline
1 & LO nonsinglet channel &  $+62.2~10^{-3}$\\ \hline
2 & NLO nonsinglet channel & $+3.24~10^{-3}$\\ \hline
3 & NNLO nonsinglet channel & $-1.08~10^{-3}$\\ \hline
4 & N$^3$LO nonsinglet channel & $-5.1~10^{-4}$\\ \hline
5 & N$^4$LO nonsinglet channel & $+1.3~10^{-5}$\\ \hline \hline
6 & $m_t$ in nonsinglet channel & $+3.7~10^{-5}$ \\ \hline
7 & $m_t$ in singlet channel & $-1.6~10^{-3}$ \\ \hline
8 & smooth thresholds in $\alpha_s$& $+1.3~10^{-3}$\\ \hline
9 & $m_b$ effects in Born term, axial channel & $-4.2~10^{-3}$ \\
\hline
10 & $m_b$ effects in LO term, vector channel & $+1.5~10^{-4}$\\
\hline\hline
11 & RS uncertainty at NLO & $-4.7~10^{-4}$\\ \hline
12 & RS uncertainty at NNLO & $-2.1~10^{-4}$\\ \hline
13 & experimental error & $\pm 5.9~10^{-3}$ \\ \hline
\end{tabular}
\end{center}
\caption{Summary of various contributions to, and uncertainties of,
$R_Z$. All numbers, except items 8,11,12 are correspond to
\bbar{MS} RS.}
\end{table}

\section{Acknowledgements}
We are grateful to our colleagues at CERN, where part of this work
was done, for their hospitality.
In its final stage the work of one of us
(A.L.K.) was supported by the Russian Fund for Fundamental Research
under Grant No. 94-02-04548-a.

\end{document}